\documentclass[preprint,preprintnumbers, prd, floatfix, superscriptaddress,nofootinbib] {revtex4-1}
\usepackage{epsfig}
\usepackage{subfigure}
\usepackage{dcolumn}
\usepackage{bm}
\usepackage[usenames ,dvipsnames]{xcolor}
\usepackage{slashed}
\usepackage{graphicx,color}
\usepackage{hyperref}
\usepackage{multirow}
\newcommand{\nn}{\nonumber}
\newcommand{\be}{\begin{eqnarray}}
\newcommand{\ee}{\end{eqnarray}}
\newcommand{\strich}[1]{#1  \! \! \slash}

\begin{document}
\title{Charmed $\Omega_c$ weak decays into $\Omega$ in the light-front quark model}

\author{Yu-Kuo Hsiao}
\email{yukuohsiao@gmail.com}
\affiliation{School of Physics and Information Engineering, Shanxi Normal University, Linfen 041004, China}

\author{Ling Yang}
\email{yangling@ihep.ac.cn}
\affiliation{School of Physics and Information Engineering, Shanxi Normal University, Linfen 041004, China}

\author{Chong-Chung Lih}
\email{cclih@phys.nthu.edu.tw}
\affiliation{Department of Optometry, Central Taiwan University of Science and Technology, Taichung 40601}

\author{Shang-Yuu Tsai}
\email{shangyuu@gmail.com}
\affiliation{School of Physics and Information Engineering, Shanxi Normal University, Linfen 041004, China}
\date{\today}

\begin{abstract}
More than ten $\Omega_c^0$ weak decay modes have been measured with 
the branching fractions relative to that of $\Omega^0_c\to\Omega^-\pi^+$.
In order to extract the absolute branching fractions,
the study of $\Omega^0_c\to\Omega^-\pi^+$ is needed.
In this work, we predict ${\cal B}_\pi\equiv
{\cal B}(\Omega_c^0\to\Omega^-\pi^+)=(5.1\pm 0.7)\times 10^{-3}$
with the $\Omega_c^0\to\Omega^-$ transition form factors calculated in the light-front quark model.
We also predict ${\cal B}_\rho\equiv
{\cal B}(\Omega_c^0\to\Omega^-\rho^+)=(14.4\pm 0.4)\times 10^{-3}$ and 
${\cal B}_e\equiv{\cal B}(\Omega_c^0\to\Omega^-e^+\nu_e)=(5.4\pm 0.2)\times 10^{-3}$.
The previous values for ${\cal B}_\rho/{\cal B}_\pi$ have been found to
deviate from the most recent observation. Nonetheless,
our ${\cal B}_\rho/{\cal B}_\pi=2.8\pm 0.4$
is able to alleviate the deviation.
Moreover, we obtain ${\cal B}_e/{\cal B}_\pi=1.1\pm 0.2$,
which is consistent with the current data.
\end{abstract}

\maketitle

\section{Introduction}
The lowest-lying singly charmed baryons include
the anti-triplet and sextet states
${\bf B}_c=(\Lambda_c^+,\Xi_c^0,\Xi_c^+)$ and
${\bf B}'_c=(\Sigma_c^{(0,+,++)},\Xi_c^{'(0,+)}, \Omega_c^0)$, respectively.
The ${\bf B}_c$ and  $\Omega_c^0$ baryons  
predominantly decay weakly~\cite{CroninHennessy:2000bz,
Ammar:2002pf,Aubert:2007bt,Yelton:2017uzv,pdg}, whereas
the $\Sigma_c$ ($\Xi'_c$) decays are strong (electromagnetic) processes.
There have been more accurate observations for the ${\bf B}_c$ weak decays
in the recent years, which have helped to improve 
the theoretical understanding of the decay processes~\cite{Lu:2016ogy,Geng:2017esc,Geng:2018plk,
Geng:2018upx,Hsiao:2019yur,Zhao:2018mov,Zou:2019kzq,Hsiao:2020iwc,Niu:2020gjw}.
With the lower production cross section of 
$\sigma(e^+e^-\to\Omega_c^0X)$~\cite{Yelton:2017uzv},
it is an uneasy task to measure $\Omega_c^0$ decays.
Consequently, most of the $\Omega_c^0$ decays 
have not been reanalysized since 1990s~\cite{AvilaAoki:1989yi,PerezMarcial:1989yh,
Singleton:1990ye,Hussain:1990ai,Korner:1992wi,Xu:1992sw,
Cheng:1993gf,Cheng:1996cs,Ivanov:1997ra},
except for those in~\cite{Pervin:2006ie,Dhir:2015tja,Zhao:2018zcb,
Gutsche:2018utw,Hu:2020nkg,Geng:2017mxn}. 

\begin{figure}[t]
\centering
\includegraphics[width=3.0 in]{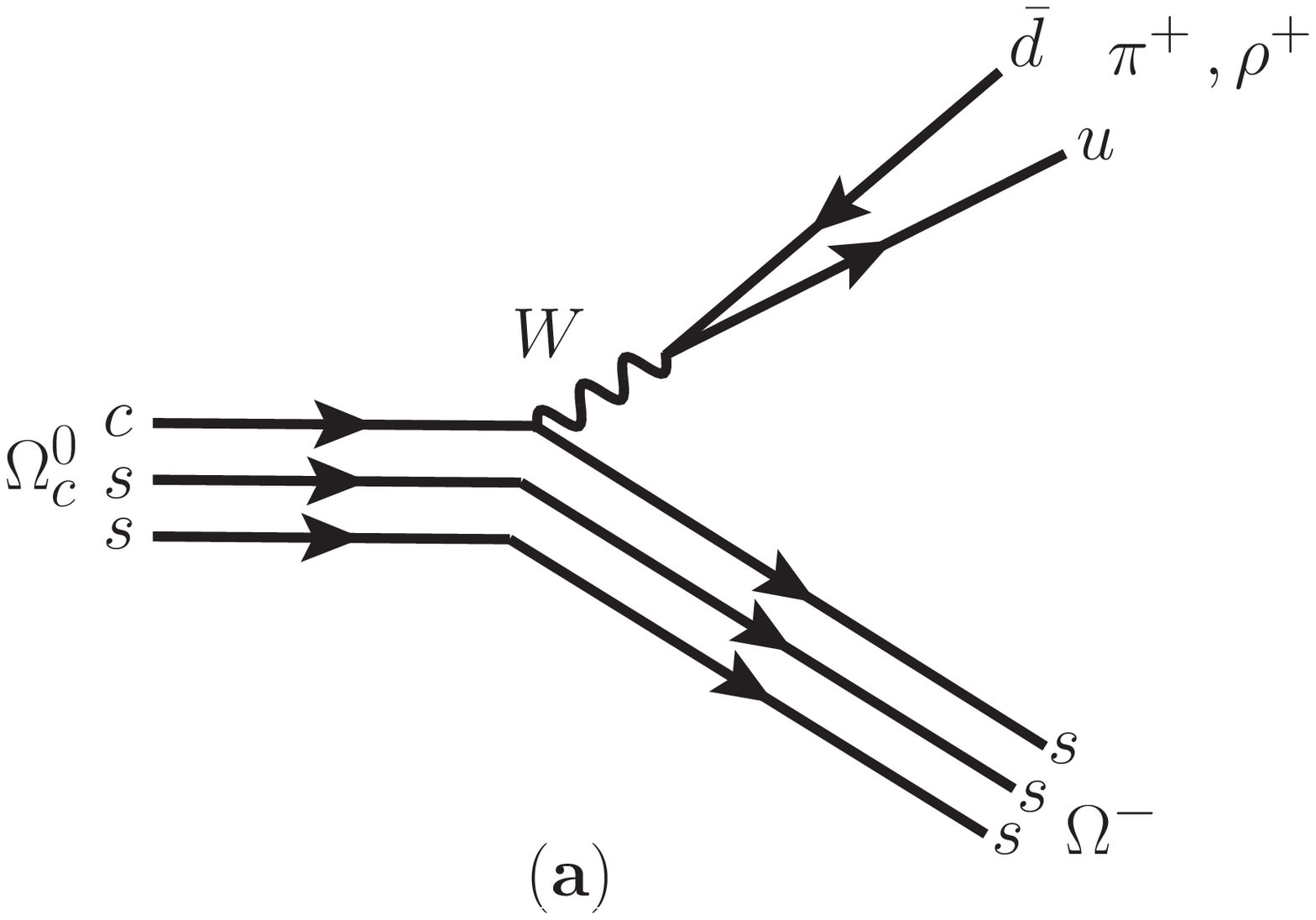}
\includegraphics[width=2.8 in]{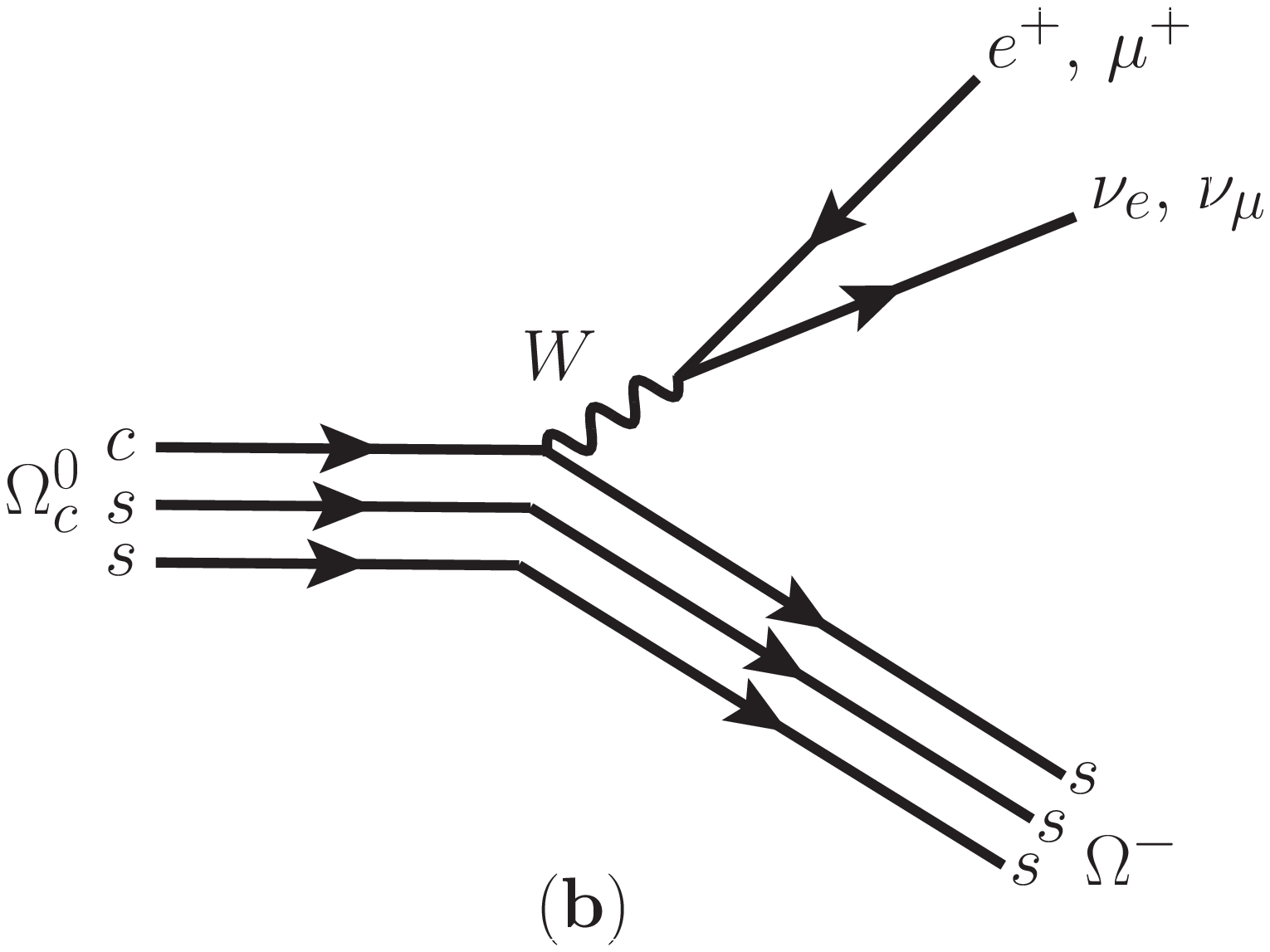}
\caption{Feynman diagrams for 
(a) $\Omega_c^0\to\Omega^-\pi^+(\rho^+)$ and 
(b) $\Omega_c^0\to\Omega^-\ell^+\nu_\ell$ with $\ell^+=e^+$ or $\mu^+$.}\label{fig:Feynman}
\end{figure}

One still manages to measure more than ten $\Omega_c^0$ decays,
such as $\Omega_c^0\to\Omega^-\rho^+$, $\Xi^0\bar K^{(*)0}$ and $\Omega^-\ell^+ \nu_\ell$, 
but with the branching fractions relative to 
${\cal B}(\Omega_c^0\to\Omega^-\pi^+)$~\cite{pdg}.
To extract the absolute branching fractions,
the study of $\Omega_c^0\to\Omega^-\pi^+$ is crucial.
Fortunately, 
the $\Omega_c^0\to\Omega^-\pi^+$ decay involves a simple topology,
which benefits its theoretical exploration.
In Fig.~\ref{fig:Feynman}a, $\Omega_c^0\to\Omega^-\pi^+$ 
is depicted to proceed through the $\Omega_c^0\to\Omega^-$ transition, 
while $\pi^+$ is produced from the external $W$-boson emission. 
Since it is a Cabibbo-allowed process
with $V_{cs}^* V_{ud}\simeq 1$, a larger branching fraction 
is promising for measurements.
Furthermore, it can be seen that
$\Omega_c^0\to\Omega^-\pi^+$ has 
a similar configuration to those of
$\Omega_c^0\to\Omega^-\rho^+$ and $\Omega_c^0\to\Omega^-\ell^+ \nu_\ell$,
as drawn in Fig.~\ref{fig:Feynman}, 
indicating that the three $\Omega_c^0$ decays are all associated 
with the $\Omega_c^0\to \Omega^-$ transition.
While $\Omega$ is a decuplet baryon 
that consists of the totally symmetric identical quarks $sss$,
behaving as a spin-3/2 particle,
the form factors of the $\Omega_c^0\to \Omega^-$ transition
can be more complicated, which hinders the calculation for the decays.
As a result, a careful investigation that relates
$\Omega_c^0\to\Omega^-\pi^+,\Omega^-\rho^+$ and $\Omega_c^0\to\Omega^-\ell^+ \nu_\ell$
has not been given yet, 
despite the fact that the topology associates them together.

Based on the quark models, 
it is possible to study the $\Omega_c^0$ decays into $\Omega^-$
with the $\Omega_c^0\to\Omega^-$ transition form factors.
However, the validity of theoretical approach needs to be tested,
which depends on if the observations, given by
\begin{eqnarray}\label{data1}
\frac{{\cal B}(\Omega_c^0\to\Omega^-\rho^+)}{{\cal B}(\Omega_c^0\to\Omega^-\pi^+)}
&=&1.7\pm 0.3\,\text{\cite{Yelton:2017uzv}}\,(>1.3\,\text{\cite{pdg}})\,, \nonumber\\
\frac{{\cal B}(\Omega_c^0\to\Omega^-e^+\nu_e)}{{\cal B}(\Omega_c^0\to\Omega^-\pi^+)}
&=&2.4\pm1.2\,\text{\cite{pdg}}\,,
\end{eqnarray}
can be interpreted.
Since the light-front quark model has been
successfully applied to the heavy hadron decays~\cite{Zhao:2018zcb,Bakker:2003up,
Ji:2000rd,Bakker:2002aw,Choi:2013ira,
Cheng:2003sm,Schlumpf:1992vq,Hsiao:2019wyd,Jaus:1991cy,
Melosh:1974cu,Dosch:1988hu,Zhao:2018mrg,Geng:2013yfa,Geng:2000if,
Ke:2012wa,Ke:2017eqo,Ke:2019smy,Hu:2020mxk}, 
in this report
we will use it to study the $\Omega_c^0\to\Omega^-$ transition form factors.
Accordingly,
we will be enabled to calculate the absolute branching fractions of 
$\Omega_c^0\to\Omega^-\pi^+(\rho^+)$ and 
$\Omega_c^0\to\Omega^- \ell^+ \nu_\ell$,
and check if the two ratios in Eq.~(\ref{data1}) can be well explained.
%
%
\section{Theoretical Framework}

\subsection{General Formalism}
To start with,
we present the effective weak Hamiltonians ${\cal H}_{H,L}$ for the
hadronic and semileptonic charmed baryon decays, respectively~\cite{Buchalla:1995vs}:
\begin{eqnarray}\label{Heff}
{\cal H}_H&=&\frac{G_F}{\sqrt 2}
V_{cs}^*V_{ud} [c_1(\bar u d)(\bar s c)+c_2(\bar s d)(\bar u c)]\,,\nonumber\\
{\cal H}_L&=&\frac{G_F}{\sqrt 2}V_{cs}^*(\bar sc)(\bar u_\nu v_\ell)\,,
\end{eqnarray}
where $G_F$ is the Fermi constant, $V_{ij}$ 
the Cabibbo-Kobayashi-Maskawa (CKM) matrix elements,
$c_{1,2}$ the effective Wilson coefficients,
$(\bar q_1 q_2)\equiv\bar q_1\gamma_\mu(1-\gamma_5)q_2$ and 
$(\bar u_\nu v_\ell)\equiv\bar u_\nu \gamma^\mu(1-\gamma_5)v_\ell$.
In terms of ${\cal H}_{H,L}$, we derive the amplitudes of 
$\Omega_c^0\to \Omega^- \pi^+(\rho^+)$ and 
$\Omega_c^0\to\Omega^- \ell^+ \nu_\ell$ as~\cite{Hsiao:2017umx,Hsiao:2018zqd}
\begin{eqnarray}\label{amp1}
{\cal M}_h\equiv{\cal M}(\Omega_c^0\to\Omega^- h^+)
&=&\frac{G_F}{\sqrt{2}}V^*_{cs}V_{ud}\,a_1
\langle\Omega^-|(\bar sc)|\Omega_c^0\rangle\langle h^+|(\bar ud)|0 \rangle\,,\nonumber\\
{\cal M}_\ell\equiv{\cal M}(\Omega_c^0\to\Omega^- \ell^+ \nu_\ell)
&=&\frac{G_F}{\sqrt{2}}V^*_{cs}
\langle\Omega^-|(\bar sc)|\Omega_c^0 \rangle(\bar u_{\nu_\ell}v_\ell)\,,
\end{eqnarray}
where $h=(\pi,\rho)$, $\ell=(e,\mu)$, 
and $a_1=c_1+c_2/N_c$ results from the factorization~\cite{Hsiao:2019ann},
with $N_c$ the color number.

With ${\bf B}'_c\,({\bf B}')$ denoting the charmed sextet (decuplet) baryon,
the matrix elements of the ${\bf B}_c'\to{\bf B}'$ transition
can be parameterized as~\cite{Zhao:2018mrg,Gutsche:2018utw}
\be
&&\langle T^\mu\rangle\equiv 
\langle {\bf B}^{\prime}(P^{\,\prime},S',S_z^\prime)|\bar q\gamma^\mu
(1-\gamma_5)c|{\bf B}'_c(P,S,S_z)\rangle \nonumber\\
&& =  \bar{u}_{\alpha}(P^{\,\prime},S_{z}^{\,\prime})
\left[\frac{P^{\alpha}}{M}\left(\gamma^{\mu}F^V_{1}
+\frac{P^{\mu}}{M} F^V_{2}
+\frac{P^{\,\prime\mu}}{M^{\prime}}F^V_{3}\right)+g^{\alpha\mu}F^V_{4}\right]
\gamma_{5}u(P,S_{z})\nonumber\\
& &\quad-\bar{u}_{\alpha}(P^{\,\prime},S_{z}^{\,\prime})
\left[\frac{P^{\alpha}}{M}\left(\gamma^{\mu}F^A_{1}
+\frac{P^{\mu} }{M}F^A_{2}
+\frac{P^{\,\prime\mu}}{M^{\prime}}F^A_{3}\right)
+g^{\alpha\mu}F^A_{4}\right]u(P,S_{z})\,,
\label{transitionVA}
\ee
where $(M,M')$ and $(S,S')=(1/2,3/2)$ represent 
the masses and spins of $({\bf B}'_c,{\bf B}')$, respectively, 
and $F^{V,A}_i$ ($i=1,2, ..,4$) the form factors to be extracted in the light-front quark model.
The matrix elements of the meson productions 
are defined as~\cite{pdg}
\begin{eqnarray}\label{decayconst}
&&
\langle\pi(p) |(\bar ud)|0 \rangle = if_\pi q^\mu\,,\nonumber\\
&&
\langle\rho(\lambda) |(\bar ud)|0 \rangle=m_\rho f_\rho \epsilon_\lambda^{\mu*}\,,
\end{eqnarray}
where $f_{\pi(\rho)}$ is the decay constant, and
$\epsilon_\lambda^\mu$ is the polarization four-vector 
with $\lambda$ denoting the helicity state.

\subsection{The light-front quark model}
The baryon bound state ${\bf B}'_{(c)}$ contains three quarks 
$q_1$, $q_2$ and $q_3$, with the subscript $c$ for $q_1=c$.
Moreover, $q_2$ and $q_3$ are combined as a diquark state $q_{[2,3]}$,
behaving as a scalar or axial-vector. Subsequently,
the baryon bound state $|{\bf B}'_{(c)}(P,S,S_z)\rangle$ in the light-front quark model
can be written as~\cite{Dosch:1988hu}
\be
|{\bf B}'_{(c)}(P,S,S_z)\rangle & = & \int\{d^{3}p_{1}\}
\{d^{3}p_{2}\}2(2\pi)^{3}\delta^{3}(\tilde{P}-\tilde{p}_{1}-\tilde{p}_{2})\nonumber \\
&  & \times\sum_{\lambda_{1},\lambda_{2}}\Psi^{SS_{z}}
(\tilde{p}_{1},\tilde{p}_{2},\lambda_{1},\lambda_{2})|q_1(p_1,\lambda_{1})q_{[2,3]}
(p_{2},\lambda_{2})\rangle\,,
\label{wf1}
\ee
where $\Psi^{SS_{z}}$ is the momentum-space wave function, and
$(p_i,\lambda_i)$ stand for momentum and helicity
of the constituent (di)quark, with $i=1,2$ for $q_1$ and $q_{[2,3]}$, respectively.
The tilde notations represent that the quantities are in the light-front frame,
and one defines $P=(P^-,P^+,P_\bot)$ and $\tilde P=(P^+,P_\bot)$,
with $P^\pm=P^0\pm P^3$ and $P_\bot=(P^1,P^2)$. 
Besides, $\tilde{p}_i$ are given by
\be
\tilde p_i=(p_i^+, p_{i\bot})~, \quad p_{i\bot}= (p_i^1, p_i^2)~,
\quad p_i^- = {m_i^2+p_{i\bot}^2\over p_i^+},
\ee with \be
&& m_1=m_{q_1}, \quad m_2=m_{q_1}+m_{q_2},\nn\\
&& p^+_1=(1-x) P^+, \quad p^+_2=x P^+,\nn\\ 
&& p_{1\bot}=(1-x) P_\bot-k_\bot, \quad p_{2\bot}=xP_\bot+k_\bot\,,
\ee 
where $x$ and $k_\perp$ are the light-front relative momentum variables
with $k_\perp$ from $\vec{k}=(k_\perp,k_z)$, 
ensuring that $P^{+}=p^+_1+p^+_2$ and $P_{\bot}=p_{1\bot}+p_{2\bot}$.
According to $e_i\equiv\sqrt{m^2_{i}+\vec{k}^2}$ and $M_0\equiv e_1+e_2$
in the Melosh transformation~\cite{Melosh:1974cu}, 
we obtain
\be
&&
x=\frac{e_2-k_z}{e_1+e_2}\,,\quad 1-x=\frac{e_1+k_z}{e_1+e_2}\,,\quad
k_z=\frac{xM_0}{2}-\frac{m^2_{2}+k^2_{\perp}}{2xM_0}\,,\nn\\
&&
M_0^2={ m_{1}^2+k_\bot^2\over 1-x}+{ m_{2}^2+k_\bot^2\over  x}\,.
\ee
Consequently, $\Psi^{SS_{z}}$ can be given 
in the following representation~\cite{Ke:2012wa,Ke:2017eqo,Zhao:2018mrg,Hu:2020mxk}:
\be
\Psi^{SS_{z}}(\tilde{p}_{1},\tilde{p}_{2},\lambda_{1},\lambda_{2})=
\frac{A^{(\prime)}}{\sqrt{2(p_{1}\cdot\bar{P}+m_{1}M_{0})}}\bar{u}(p_{1},\lambda_{1})
\Gamma_{S,A}^{(\alpha)} u(\bar{P},S_{z})\phi(x,k_{\perp})\,,
\label{wf2}
\ee
with 
\be
&&
A=\sqrt{\frac{3(m_{1}M_{0}+p_{1}\cdot\bar{P})}{3m_{1}M_{0}+p_{1}\cdot\bar{P}+
2(p_{1}\cdot p_{2})(p_{2}\cdot\bar{P})/m_{2}^{2}}}\,,\nonumber\\
&&
\Gamma_S=1,\;\;
\Gamma_{A}=-\frac{1}{\sqrt{3}}\gamma_{5} \strich\epsilon^{*}(p_{2},\lambda_{2})\,,\nonumber
\ee
and
\be
&&
A'=\sqrt{\frac{3m_{2}^{2}M_{0}^{2}}{2m_{2}^{2}M_{0}^{2}+(p_{2}\cdot\bar{P})^{2}}}\,,\;\;
\Gamma_{A}^{\alpha}=\epsilon^{*\alpha}(p_{2},\lambda_{2})\,,
\label{wf3}
\ee
where the vertex function $\Gamma_{S(A)}$ 
is for the scalar (axial-vector) diquark in ${\bf B}'_c$, and 
$\Gamma_A^\alpha$ for the axial-vector diquark in ${\bf B}'$.
We have used the variable $\bar P\equiv p_1+p_2$ 
to describe the internal motions of the constituent quarks in the baryon~\cite{Jaus:1991cy},
which leads to $(\bar P_\mu \gamma^\mu-M_0)u(\bar{P},S_{z})=0$,
different from $(P_\mu \gamma^\mu-M)u(P,S_{z})=0$.
For the momentum distribution, 
$\phi(x,k_{\perp})$ is presented as the Gaussian-type wave function,
given by
\be
\phi(x,k_{\perp})=4\left(\frac{\pi}{\beta^{2}}\right)^{3/4}\sqrt{\frac{e_{1}e_{2}}{x(1-x)M_{0}}}\exp
\left(\frac{-\vec{k}^{2}}{2\beta^{2}}\right)\,,
\label{wf4}
\ee
where $\beta$ shapes the distribution.

Using $|{\bf B}'_c(P,S,S_z)\rangle$ and $|{\bf B}'(P,'S',S'_z)\rangle$
from Eq.~(\ref{wf1}) and
their components in Eqs.~(\ref{wf2}), (\ref{wf3}) and (\ref{wf4}),
we derive the matrix elements of 
the ${\bf B}'_c\to{\bf B}'$ transition in Eq.~(\ref{transitionVA}) as
\be
&& 
\langle \bar T^\mu\rangle\equiv 
\langle {\bf B}^{\prime}(P^{\,\prime},S',S_z^\prime)|\bar q\gamma^\mu
(1-\gamma_5)c|{\bf B}'_c(P,S,S_z)\rangle\nonumber \\
&&~\, =  \int\{d^{3}p_{2}\}\frac{\phi^{\prime}(x^{\prime},k_{\perp}^{\prime})\phi(x,k_{\perp})}
{2\sqrt{p_{1}^{+}p_{1}^{\prime+}(p_{1}\cdot\bar{P}+m_{1}M_{0})(p_{1}^{\prime}\cdot\bar{P}^{\,\prime}
+m_{1}^{\prime}M_{0}^{\prime})}}\nonumber \\
&& 
\times\sum_{\lambda_{2}}\bar{u}_{\alpha}(\bar{P}^{\,\prime},S_{z}^{\,\prime})
\left[\bar{\Gamma}^{\,\prime\alpha}_{A}(\strich p_{1}^{\prime}+m_{1}^{\prime})
\gamma^{\mu}(1-\gamma_{5})(\strich p_{1}+m_{1})\Gamma_{A}\right]u(\bar{P},S_{z})\,,
\label{transitionVA2}
\ee
with $m_1=m_c$, $m'_1=m_q$ and $\bar \Gamma=\gamma^0 \Gamma^\dagger\gamma^0$.
We define
$J_{5\,j}^{\mu}=\bar{u}(\Gamma_{5}^{\mu\beta})_{j}u_{\beta}$ and
$\bar J_{5\,j}^{\mu}=\bar{u}(\bar \Gamma_{5}^{\mu\beta})_{j}u_{\beta}$ with $j=1,2,...,4$,
where
\be
&&(\Gamma_{5}^{\mu\beta})_j=
\{\gamma^{\mu}P^{\beta},P^{\,\prime\mu}P^{\beta},P^{\mu}P^{\beta},g^{\mu\beta}\}\gamma_{5}\,,\nn\\
&&(\bar{\Gamma}_{5}^{\mu\beta})_j=
\{\gamma^{\mu}\bar{P}^{\beta},\bar{P}^{\,\prime\mu}\bar{P}^{\beta},\bar{P}^{\mu}\bar{P}^{\beta},g^{\mu\beta}\}\gamma_{5}\,.
\ee
Then, we multiply $J_{5\,j}$ ($\bar J_{5\,j}$) by $\langle T\rangle$ ($\langle \bar T\rangle$)
as $F_{5\,j}\equiv J_{5\,j}\cdot \langle T\rangle$ and 
$\bar F_{5\,j}\equiv \bar J_{5\,j}\cdot \langle \bar T\rangle$ with 
$\langle T\rangle$ and $\langle \bar T\rangle$ 
in Eqs.~(\ref{transitionVA}) and (\ref{transitionVA2}), respectively,
resulting in~\cite{Zhao:2018mrg}
\be\label{F5j}
&&
F_{5\,j}=Tr\bigg\{
u_{\beta}\bar{u}_{\alpha}
\left[\frac{P^{\alpha}}{M}\left(\gamma^{\mu}F^V_{1}
+\frac{P^{\mu}}{M} F^V_{2}
+\frac{P^{\,\prime\mu}}{M^{\prime}}F^V_{3}\right)+g^{\alpha\mu}F^V_{4}\right]
\gamma_{5}\bar u({\Gamma}_{5\mu}^{\beta})_j\bigg\}\,,\nn\\
&&
\bar F_{5\,j}= \int\{d^{3}p_{2}\}\frac{\phi^{\prime}(x^{\prime},k_{\perp}^{\prime})\phi(x,k_{\perp})}
{2\sqrt{p_{1}^{+}p_{1}^{\prime+}(p_{1}\cdot\bar{P}+m_{1}M_{0})(p_{1}^{\prime}\cdot\bar{P}^{\,\prime}
+m_{1}^{\prime}M_{0}^{\prime})}}
\nn\\
&&
\times\sum_{\lambda_{2}}
Tr\bigg\{u_{\beta}\bar{u}_{\alpha}
\left[\bar{\Gamma}^{\,\prime\alpha}_{A}(\strich p_{1}^{\prime}+m_{1}^{\prime})
\gamma^{\mu}(\strich p_{1}+m_{1})\Gamma_{A}\right]
 u(\bar{\Gamma}_{5\mu}^{\beta})_j\bigg\}\,.
\ee
In the connection of $F_{5\,j}=\bar F_{5\,j}$,
we construct four equations. By solving the four equations,
the four form factors $F^V_1$, $F^V_2$, $F^V_3$ and $F^V_4$ can be extracted.
The form factors $F^A_i$ can be obtained in the same way. 

\subsection{Branching fractions in the helicity basis}
One can present the amplitude of $\Omega_c^0\to\Omega^- h^+(\Omega^- \ell^+\nu_\ell)$ 
in the helicity basis of $H_{\lambda_\Omega \lambda_{h(\ell)}}$~\cite{Gutsche:2018utw,Zhao:2018mrg},
where $\lambda_\Omega=\pm 3/2,\pm 1/2$ 
represent the helicity states of the $\Omega^-$ baryon, and 
$\lambda_{h,\ell}$ those of $h^+$ and $\ell^+\nu_\ell$. 
Substituting the matrix elements in Eqs.~(\ref{amp1}) 
with those in Eqs.~(\ref{transitionVA}) and (\ref{decayconst}), 
the amplitudes in the helicity basis now read
$\sqrt 2{\cal M}_h=
(i)\sum_{\lambda_\Omega,\lambda_h}G_F V^*_{cs}V_{ud}\,a_1 m_h f_h H_{\lambda_\Omega \lambda_h}$ and 
$\sqrt 2{\cal M}_\ell=\sum_{\lambda_\Omega,\lambda_\ell}G_F V^*_{cs} H_{\lambda_\Omega \lambda_\ell}$, 
where $H_{\lambda_\Omega\lambda_f}=H^V_{\lambda_\Omega \lambda_f}-H^A_{\lambda_\Omega \lambda_f}$
with $f=(h,\ell)$. Explicitly, 
$H^{V(A)}_{\lambda_\Omega \lambda_f}$ is written as~\cite{Gutsche:2018utw}
\be
H^{V(A)}_{\lambda_\Omega \lambda_f}
\equiv\langle\Omega^-|\bar s\gamma_\mu(\gamma_5)c|\Omega_c^0\rangle
\varepsilon^\mu_f\,,
\ee
with $\varepsilon^\mu_h=(q^\mu/\sqrt{q^2},\epsilon_\lambda^{\mu*})$
for $h=(\pi,\rho)$. For the semi-leptonic decay, 
since the $\ell^+\nu_\ell$ system behaves as a scalar or vector,
$\varepsilon^\mu_\ell=q^\mu/\sqrt{q^2}$ or $\epsilon_\lambda^{\mu\,*}$.
The $\pi$ meson only has a zero helicity state, denoted by $\lambda_\pi=\bar 0$.
On the other hand, the three helicity states of $\rho$ are denoted by $\lambda_\rho=(1,0,-1)$.
For the lepton pair, we assign $\lambda_\ell=\lambda_\pi$ or $\lambda_\rho$.
Subsequently, 
we expand $H^{V(A)}_{\lambda_\Omega \lambda_f}$ as
\begin{eqnarray}\label{Hpi}
H_{\frac12 {\bar 0}}^{V(A)} &=&\sqrt{\frac{2}{3}\frac{Q^2_{\pm}}{q^2}}
\left(\frac{Q^2_\mp}{2MM'}\right)
(F_1^{V(A)} M_\pm \mp  F_2^{V(A)}\bar M_+ \mp  F_3^{V(A)}\bar M'_- \mp F_4^{V(A)} M )\,,
\end{eqnarray}
for $\varepsilon^\mu_f=q^\mu/\sqrt{q^2}$,
where $M_\pm = M\pm M'$, $Q^2_\pm = M_\pm^2 - q^2$, 
and $\bar M_{\pm}^{(\prime)}=(M_+M_-\pm q^2)/(2M^{(\prime)})$.
We also obtain
\begin{eqnarray}\label{Hrho}
&&
H_{\frac321}^{V(A)} = \mp \sqrt{Q^2_\mp} \, F_4^{V(A)}\,,\nonumber\\
&&
H_{\frac121}^{V(A)}=-\sqrt{\frac{Q^2_\mp}{3}}
\left[F_1^{V(A)} \left(\frac{Q^2_\pm}{M M'}\right) -F_4^{V(A)}\right]\,,\nonumber\\
&&
H_{\frac120}^{V(A)}= \sqrt{\frac{2}{3}\frac{Q^2_\mp}{q^2}}
\left[ F_1^{V(A)} \left(\frac{Q^2_\pm M_\mp}{2MM'}\right)
\mp\left(F_2^{V(A)}+F_3^{V(A)}\frac{M}{M'}\right) \left(\frac{|\vec{P}'|^2}{M'}\right) 
\mp F_4^{V(A)}\bar M'_- \right]\,,
\end{eqnarray}
for $\varepsilon^\mu_f=\epsilon_\lambda^{\mu*}$,
with $|\vec{P}'|=\sqrt{Q^2_+ Q^2_-}/(2M)$. 
Note that the expansions in Eqs.~(\ref{Hpi}) and (\ref{Hrho}) have satisfied
$\lambda_{\Omega_c}=\lambda_\Omega-\lambda_f$ for the helicity conservation,
with $\lambda_{\Omega_c}=\pm 1/2$. The branching fractions then read
\begin{eqnarray}
{\cal B}_h\equiv {\cal B}(\Omega^0_c\to\Omega^- h^+)&=&
\frac{\tau_{\Omega_c}G_F^2|\vec{P}'|}{32\pi m_{\Omega_c}^2}
|V_{cs}V_{ud}^*|^2\,a_1^2 m_h^2 f_h^2 H_h^2\,,\nonumber\\
{\cal B}_\ell\equiv{\cal B}(\Omega_c^0\to\Omega^- \ell^+\nu_\ell)
&=&\frac{\tau_{\Omega_c}G_F^2|V_{cs}|^2}{192\pi^3 m_{\Omega_c}^2}
\int^{(m_{\Omega_c}-m_\Omega)^2}_{m_\ell^2}dq^2
\left(\frac{|\vec{P}'|(q^2-m_\ell^2)^2}{q^2}\right)H_\ell^2\,,
\end{eqnarray}
where
\begin{eqnarray}\label{3H}
H_\pi^2&=&
\left|H_{\frac12{\bar 0}}\right|^2+\left|H_{{-\frac12}{\bar 0}}\right|^2\,,\nonumber\\
H_\rho^2&=&
\left|H_{\frac321}\right|^2+\left|H_{\frac121}\right|^2+\left|H_{\frac120}\right|^2
+\left|H_{-\frac120}\right|^2+\left|H_{-\frac12-1}\right|^2+\left|H_{-\frac32-1}\right|^2\,,\nonumber\\
H_\ell^2&=&\left(1+\frac{m_\ell^2}{2q^2}\right)H_\rho^2+\frac{3m_\ell^2}{2q^2}H_\pi^2\,,
\end{eqnarray}
with $\tau_{\Omega_c}$ the $\Omega_c^0$ lifetime.

\section{Numerical analysis}
In the Wolfenstein parameterization,
the CKM matrix elements 
are adopted as $V_{cs}=V_{ud}=1-\lambda^2/2$ with $\lambda=0.22453\pm 0.00044$~\cite{pdg}.
We take the lifetime and mass of the $\Omega_c^0$ baryon
and the decay constants $(f_\pi,f_\rho)=(132,216)$~MeV from the PDG~\cite{pdg}. 
With $(c_1,c_2)=(1.26,-0.51)$ at the $m_c$ scale~\cite{Buchalla:1995vs},
we determine $a_1$. In the generalized factorization, 
$N_c$ is taken as an effective color number with 
$N_c=(2,3,\infty)$~\cite{Hu:2020nkg,Gutsche:2018utw,Hsiao:2019wyd,Hsiao:2019ann},
in order to estimate the non-factorizable effects.
For the $\Omega_c^+(css)\to \Omega^-(sss)$ transition form factors,
the theoretical inputs of the quark masses and parameter $\beta$ in Eq.~(\ref{F5j})
are given by~\cite{Geng:2013yfa,Geng:2000if}
\be\label{qmass}
&&
m_{1}=m_{c}=(1.35\pm0.05)~\mathrm{GeV}\,,\quad m_{1}^{\prime}=m_{s}=0.38~\mathrm{GeV}\,,\quad 
m_{2}=2m_s=0.76~\mathrm{GeV}\,,\nonumber\\&&
\beta_c=0.60~\mathrm{GeV}\,,
\quad \beta_s=0.46~\mathrm{GeV}\,,
\ee
where $\beta_{c(s)}$ is to determine $\phi^{(\prime)}(x^{(\prime)},k_{\perp}^{(\prime)})$
for $\Omega_c^0$ $(\Omega^-)$. We hence extract $F^V_i$ and $F^A_i$ in Table~\ref{ffactor}.
For the momentum dependence, we have used
the double-pole parameterization:
\begin{equation}
F(q^2)=\frac{F(0)}{1-a\left(q^2/m_F^2\right)+b\left(q^4/m_F^4\right)}\,,
\label{eq:LFparameters}
\end{equation}
with $m_F=1.86$~GeV. 
%
\begin{table}[t!]
\caption{The $\Omega^0_c\to\Omega^-$ transition form factors with $F(0)$ at $q^2=0$, 
where $\delta\equiv \delta m_c/m_c=\pm 0.04$ from Eq.~(\ref{qmass}).}\label{tab1}
\vskip 0.2in
\label{ffactor}
{\scriptsize
\begin{tabular}{|c|c|r|r|} \hline
    & $F(0)$ & $a\;\;\;$ & $b\;\;\;$
\\ \hline \hline
$F^V_{1}$ & $0.54+0.13\delta$ & $-0.27$ & $ 1.65$
\\ 
$F^V_{2}$ & $0.35-0.36\delta$ & $-30.00$ & $96.82$
\\ 
$F^V_{3}$ & $0.33+0.59\delta$ & $0.96$ & $9.25$
\\ 
$F^V_{4}$ & $0.97+0.22\delta$ & $-0.53$ & $1.41$
\\ \hline
\end{tabular}
\begin{tabular}{|c|c|r|r|} \hline
& $F(0)$ & $a\;\;\;$ & $b\;\;\;$
\\ \hline \hline
$F^A_{1}$ & $\;\;\;2.05+1.38\delta$ & $-3.66$ & $1.41$
\\ 
$F^A_{2}$ & $-0.06+0.33\delta$ & $-1.15$ & $71.66$
\\ 
$F^A_{3}$ & $-1.32-0.32\delta$ & $-4.01$ & $5.68$
\\ 
$F^A_{4}$ & $-0.44+0.11\delta$ & $-1.29$ & $-0.58$
\\ \hline
\end{tabular}}
\end{table}
Using the theoretical inputs, we calculate the branching fractions, 
whose results are given in Table~\ref{tab2}.

\section{Discussions and Conclusions}
\begin{table}[t!]
\caption{Branching fractions of (non-)leptonic $\Omega_c^0$ decays
and their ratios, where ${\cal R}_{\rho(e)/\pi}\equiv {\cal B}_{\rho(e)}/{\cal B}_\pi$.
The three numbers in the parenthesis correspond to $N_c=(2,3,\infty)$, and
the errors come from the uncertainties of the form factors in Table~\ref{ffactor}.}\label{tab2}
\begin{center}
{\scriptsize
\begin{tabular}{|l||c|cccc|c|}
\hline
${\cal B}({\cal R})$
&our work 
&Ref.~\cite{Xu:1992sw}
&Ref.~\cite{Cheng:1996cs}
&Ref.~\cite{Gutsche:2018utw} 
&Ref.~\cite{Pervin:2006ie}
&data~\cite{pdg,Yelton:2017uzv}\\
\hline
\hline
$10^3{\cal B}_\pi$ 
&$(5.1\pm 0.7,6.0\pm 0.8,8.0\pm 1.0)$
&$(56.6,66.5,88.9)$
&$(36.0,42.3,56.6)$
&$(-,-,2)$
&
&\\
$10^3{\cal B}_\rho$ 
&$(14.4\pm 0.4,17.0\pm 0.5,22.1\pm 0.6)$
&$(307.0,361.1,482.5)$
&$(126.7,149.0,199.1)$
&$(-,-,19)$
&
&\\\hline
$10^3{\cal B}_{e}$
&$5.4\pm 0.2$
&
&
&
&127
&\\
$10^3{\cal B}_{\mu}$
&$5.0\pm 0.2$
&
&
&
&
&\\\hline
${\cal R}_{\rho/\pi}$
&$2.8\pm 0.4$
&5.4
&3.5
&9.5
&
&$1.7\pm 0.3$ ($>1.3$)\\
${\cal R}_{e/\pi}$
&$(1.1\pm 0.2,0.9\pm 0.1,0.7\pm 0.1)$
&
&
&
&
&$2.4\pm1.2$\\\hline
\end{tabular}}
\end{center}
\end{table}
%
In Table~\ref{tab2}, we present ${\cal B}_{\pi}$ and ${\cal B}_{\rho}$ with $N_c=(2,3,\infty)$.
The errors come from the form factors in Table~\ref{tab1},
of which the uncertainties are correlated with the charm quark mass.
By comparison,
${\cal B}_\pi$ and ${\cal B}_\rho$ are compatible with the values in Ref.~\cite{Gutsche:2018utw};
however, an order of magnitude smaller than those in Refs.~\cite{Xu:1992sw,Cheng:1996cs},
whose values are obtained with the total decay widths 
$\Gamma_{\pi(\rho)}=2.09 a_1^2(11.34 a_1^2)\times 10^{11}$~s$^{-1}$ and 
$\Gamma_{\pi(\rho)}=1.33 a_1^2(4.68 a_1^2)\times 10^{11}$~s$^{-1}$, respectively.
We also predict 
${\cal B}_e=(5.4\pm 0.2)\times 10^{-3}$ as well as ${\cal B}_\mu\simeq {\cal B}_e$,
which is much smaller than the value of $127\times 10^{-3}$ in~\cite{Pervin:2006ie}.
Only the ratios ${\cal R}_{\rho/\pi}$ and ${\cal R}_{e/\pi}$
have been actually observed so far. In our work,
${\cal R}_{\rho/\pi}=2.8\pm 0.4$ is able to alleviate the inconsistency 
between the previous value and the most recent observation.
We obtain ${\cal R}_{e/\pi}=1.1\pm 0.2$ with $N_c=2$
to be consistent with the data, which indicates
that $({\cal B}_\pi,{\cal B}_\rho)=(5.1\pm 0.7,14.4\pm 0.4)\times 10^{-3}$
with $N_c=2$ are more favorable.

The helicity amplitudes can be used to better understand how the form factors contribute 
to the branching fractions. 
With the identity $H^{V(A)}_{-\lambda_\Omega -\lambda_f} = \mp H^{V(A)}_{\lambda_\Omega\lambda_f}$
for the ${\bf B}'_c(J^P=1/2^+)$ to ${\bf B}'(J^P=3/2^+)$ transition~\cite{Gutsche:2018utw}, 
$H_\pi^2$ in Eq.~(\ref{3H}) can be rewritten as
$H_\pi^2=2(|H_{\frac12 {\bar 0}}^V|^2 +|H_{\frac12 {\bar 0}}^A|^2)$.
From the pre-factors in Eq.~(\ref{Hpi}), we estimate the ratio of
$|H_{\frac12 {\bar 0}}^V|^2/|H_{\frac12 {\bar 0}}^A|^2\simeq 0.05$,
which shows that $H_{\frac12 {\bar 0}}^A$ dominates ${\cal B}_\pi$, 
instead of $H_{\frac12 {\bar 0}}^V$. More specifically, 
it is the $F_4^A$ term in $H_{\frac12 {\bar 0}}^A$ that gives
the main contribution to the branching fraction.
By contrast, the $F_{1,3}^A$ terms in $H_{\frac12 {\bar 0}}^A$ largely cancel each other,
which is caused by $F_1^A M_- \simeq F_3^A \bar M'_-$
and a minus sign between $F_1^A$ and $F_3^A$ (see Table~\ref{tab1}); besides,
the $F_2^A$ term with a small $F_2^A(0)$ is ignorable.

Likewise, we obtain $H_\rho^2=2(|H_\rho^V|^2+|H_\rho^A|^2)$ for ${\cal B}_\rho$,
where $|H_\rho^{V(A)}|^2=
|H_{\frac321}^{V(A)}|^2+|H_{\frac121}^{V(A)}|^2+|H_{\frac120}^{V(A)}|^2$.
We find that $|H_\rho^A|^2$ is ten times larger than $|H_\rho^V|^2$.
Moreover, $H_{\frac120}^A$ 
is similar to $H_{\frac12\bar 0}^A$, 
where the $F_{1,3}^A$ terms largely cancel each other, $F_2^A$ is ignorable,
and $F_4^A$ gives the main contribution. While 
$F_1^A$ and $F_4^A$ in $H_{\frac121}^A$ have a positive interference,
giving 20\% of ${\cal B}_\rho$, $F_4^A$ in $H_{\frac321}^A$ singly contributes 35\%.
In Eq.~(\ref{3H}), the factor of $m_\ell^2/q^2$ with $m_\ell\simeq 0$
should be much suppressed, such that $H_\ell^2\simeq H_\rho^2$.
Therefore, ${\cal B}_\ell$ receives
the main contributions from the $F_4^A$ terms 
in $H_{\frac120}^A$, $H_{\frac121}^A$ and $H_{\frac321}^A$,
which is similar to the analysis for ${\cal B}_\rho$.

\newpage
In summary,
we have studied the $\Omega^0_c\to\Omega^-\pi^+,\Omega^-\rho^+$ and 
$\Omega^0_c\to\Omega^-\ell^+\nu_\ell$ decays, 
which proceed through the $\Omega_c^0\to\Omega^-$ transition
and the formation of the meson $\pi^+(\rho^+)$ or lepton pair from
the external $W$-boson emission.
With the form factors of the $\Omega_c^0\to\Omega^-$ transition,
calculated in the light-front quark model, we have predicted 
${\cal B}(\Omega_c^0\to \Omega^-\pi^+,\Omega^-\rho^+)=(5.1\pm 0.7,14.4\pm 0.4)\times 10^{-3}$
and ${\cal B}(\Omega_c^0\to\Omega^- e^+\nu_e)=(5.4\pm 0.2)\times 10^{-3}$.
While the previous studies have given the ${\cal R}_{\rho/\pi}$ values
deviating from the most recent observation, we have presented
${\cal R}_{\rho/\pi}=2.8\pm 0.4$ to alleviate the deviation.
Moreover, we have obtained ${\cal R}_{e/\pi}=1.1\pm 0.2$,
consistent with the current data.

\section*{ACKNOWLEDGMENTS}
YKH was supported in part by National Science Foundation of China (No. 11675030).
CCL was supported in part by CTUST (No. CTU109-P-108).


\begin{thebibliography}{99}

\bibitem{CroninHennessy:2000bz}
D.~Cronin-Hennessy \textit{et al.} (CLEO Collaboration),
Phys. Rev. Lett. {\bf 86}, 3730 (2001). 

\bibitem{Ammar:2002pf}
R.~Ammar \textit{et al.} (CLEO Collaboration),
Phys. Rev. Lett. {\bf 89}, 171803 (2002). 

\bibitem{Aubert:2007bt}
B.~Aubert \textit{et al.} (BaBar Collaboration),
Phys. Rev. Lett. {\bf 99}, 062001 (2007). 

\bibitem{Yelton:2017uzv}
J.~Yelton \textit{et al.} (Belle Collaboration),
Phys. Rev. D {\bf 97}, 032001 (2018). 

\bibitem{pdg}
M.~Tanabashi {\it et al.} [Particle Data Group], 
Phys.\ Rev.\ D {\bf 98}, 030001 (2018).

\bibitem{Lu:2016ogy}
C.D.~Lu, W.~Wang and F.S.~Yu,
Phys.\ Rev.\ D {\bf 93}, 056008 (2016). 

\bibitem{Geng:2017esc}
C.Q.~Geng, Y.K.~Hsiao, Y.H.~Lin and L.L. Liu,
Phys.\ Lett.\ B {\bf 776}, 265 (2017). 

\bibitem{Geng:2018plk}
C.Q.~Geng, Y.K.~Hsiao, C.W.~Liu and T.H.~Tsai,
Phys.\ Rev.\ D {\bf 97}, 073006 (2018). 

\bibitem{Geng:2018upx}
C.Q.~Geng, Y.K.~Hsiao, C.W.~Liu and T.H.~Tsai,
Phys.\ Rev.\ D {\bf 99}, 073003 (2019). 

\bibitem{Hsiao:2019yur}
Y.K.~Hsiao, Y.~Yu and H.J.~Zhao,
Phys.\ Lett.\ B {\bf 792}, 35 (2019). 

\bibitem{Zhao:2018mov} 
H.J.~Zhao, Y.L.~Wang, Y.K.~Hsiao and Y.~Yu,
JHEP {\bf 2002}, 165 (2020). 

\bibitem{Zou:2019kzq} 
J.~Zou, F.~Xu, G.~Meng and H.Y.~Cheng,
Phys.\ Rev.\ D {\bf 101}, 014011 (2020). 

\bibitem{Hsiao:2020iwc} 
Y.K.~Hsiao, Q.~Yi, S.T.~Cai and H.J.~Zhao,
arXiv:2006.15291 [hep-ph].

\bibitem{Niu:2020gjw} 
P.Y.~Niu, J.M.~Richard, Q.~Wang and Q.~Zhao,
arXiv:2003.09323 [hep-ph].

\bibitem{AvilaAoki:1989yi}
M.~Avila-Aoki, A.~Garcia, R.~Huerta and R.~Perez-Marcial,
Phys. Rev. D {\bf 40}, 2944 (1989).

\bibitem{PerezMarcial:1989yh}
R.~Perez-Marcial, R.~Huerta, A.~Garcia and M.~Avila-Aoki,
Phys. Rev. D {\bf 40}, 2955 (1989).

\bibitem{Singleton:1990ye}
R.L.~Singleton, 
Phys. Rev. D {\bf 43}, 2939 (1991).

\bibitem{Hussain:1990ai}
F.~Hussain and J.~Korner,
Z. Phys. C {\bf 51}, 607 (1991).

\bibitem{Korner:1992wi}
J.~Korner and M.~Kramer,
Z. Phys. C {\bf 55}, 659 (1992).

\bibitem{Xu:1992sw}
Q.~Xu and A.~Kamal,
Phys. Rev. D {\bf 46}, 3836 (1992).

\bibitem{Cheng:1993gf}
H.Y.~Cheng and B.~Tseng,
Phys. Rev. D {\bf 48}, 4188 (1993). 

\bibitem{Cheng:1996cs}
H.Y.~Cheng,
Phys. Rev. D {\bf 56}, 2799 (1997). 


\bibitem{Ivanov:1997ra}
M.A.~Ivanov, J.~Korner, V.E.~Lyubovitskij and A.~Rusetsky,
Phys. Rev. D {\bf 57}, 5632 (1998). 

\bibitem{Pervin:2006ie}
M.~Pervin, W.~Roberts and S.~Capstick,
Phys. Rev. C {\bf 74}, 025205 (2006). 

\bibitem{Dhir:2015tja}
R.~Dhir and C.~Kim, 
Phys. Rev. D {\bf 91}, 114008 (2015). 

\bibitem{Geng:2017mxn}
C.Q.~Geng, Y.K.~Hsiao, C.W.~Liu and T.H.~Tsai,
JHEP {\bf 1711}, 147 (2017). 

\bibitem{Zhao:2018zcb}
Z.X.~Zhao, 
Chin. Phys. C {\bf 42}, 093101 (2018). 

\bibitem{Gutsche:2018utw}
T.~Gutsche, M.A.~Ivanov, J.G.~Korner and V.E.~Lyubovitskij,
Phys. Rev. D {\bf 98}, 074011 (2018). 

\bibitem{Hu:2020nkg}
S.~Hu, G.~Meng and F.~Xu, 
Phys. Rev. D {\bf 101}, 094033 (2020). 





\bibitem{Melosh:1974cu}
H.J.~Melosh, 
Phys. Rev. D {\bf 9}, 1095 (1974).

\bibitem{Dosch:1988hu}
H.G.~Dosch, M.~Jamin and B.~Stech, 
Z. Phys. C {\bf 42}, 167 (1989).

\bibitem{Jaus:1991cy} 
W.~Jaus, 
Phys.\ Rev.\ D {\bf 44}, 2851 (1991).

\bibitem{Schlumpf:1992vq}
F.~Schlumpf, 
Phys.\ Rev.\ D {\bf 47}, 4114 (1993);
Erratum: [Phys.\ Rev.\ D {\bf 49}, 6246 (1994)]. 

\bibitem{Geng:2000if} 
C.Q.~Geng, C.C.~Lih and W.M.~Zhang,
Mod.\ Phys.\ Lett.\ A {\bf 15}, 2087 (2000). 

\bibitem{Ji:2000rd}
C.R.~Ji and C.~Mitchell, 
Phys.\ Rev.\ D {\bf 62}, 085020 (2000).

\bibitem{Bakker:2002aw}
B.L.G.~Bakker and C.R.~Ji, 
Phys.\ Rev.\ D {\bf 65}, 073002 (2002). 

\bibitem{Bakker:2003up}
B.L.G.~Bakker, H.M.~Choi and C.R.~Ji,
Phys.\ Rev.\ D {\bf 67}, 113007 (2003). 

\bibitem{Cheng:2003sm}
H.Y.~Cheng, C.K.~Chua and C.W.~Hwang,
Phys.\ Rev.\ D {\bf 69}, 074025 (2004). 

\bibitem{Choi:2013ira}
H.M.~Choi and C.R.~Ji, 
Few Body Syst.\  {\bf 55}, 435 (2014). 

\bibitem{Geng:2013yfa} 
C.Q.~Geng and C.C.~Lih,
Eur.\ Phys.\ J.\ C {\bf 73}, 2505 (2013). 


\bibitem{Ke:2012wa}
H.W.~Ke, X.H.~Yuan, X.Q.~Li, Z.T.~Wei and Y.X.~Zhang,
Phys. Rev. D {\bf 86}, 114005 (2012). 

\bibitem{Ke:2017eqo}
H.W.~Ke, N.~Hao and X.Q.~Li,
J. Phys. G {\bf46}, 115003 (2019). 

\bibitem{Zhao:2018mrg}
Z.X.~Zhao, 
Eur. Phys. J. C {\bf 78}, 756 (2018). 

\bibitem{Hu:2020mxk} 
X.H.~Hu, R.H.~Li and Z.P.~Xing,
Eur.\ Phys.\ J.\ C {\bf 80}, 320 (2020). 

\bibitem{Ke:2019smy}
H.W.~Ke, N.~Hao and X.Q.~Li,
Eur. Phys. J. C {\bf 79}, 540 (2019). 



\bibitem{Hsiao:2019wyd} 
Y.K.~Hsiao, S.Y.~Tsai, C.C.~Lih and E.~Rodrigues,
JHEP {\bf 2004}, 035 (2020). 
\bibitem{Buchalla:1995vs}
G.~Buchalla, A.~J.~Buras and M.~E.~Lautenbacher,
Rev. Mod. Phys. {\bf 68}, 1125 (1996). 

\bibitem{Hsiao:2017umx} 
Y.K.~Hsiao and C.Q.~Geng,
Eur.\ Phys.\ J.\ C {\bf 77}, 714 (2017). 

\bibitem{Hsiao:2018zqd} 
Y.K.~Hsiao and C.Q.~Geng,
Phys.\ Lett.\ B {\bf 782}, 728 (2018). 

\bibitem{Hsiao:2019ann} 
Y.K.~Hsiao, S.Y.~Tsai and E.~Rodrigues,
Eur.\ Phys.\ J.\ C {\bf 80}, 565 (2020). 

\end{thebibliography}
\end{document}